\documentclass{elsarticle}

\usepackage{lineno,hyperref}

\journal{Nuclear Instruments and Methods in Physics Research A}

\begin{document}

\begin{frontmatter}

\title{On the initial approximation of charged particle tracks in detectors with linear sensing elements}

\author[1]{A.V.~Belyaev}
\ead{alxbljv@yandex.ru}
\author[1]{\fbox{S.A.~Avramenko}}
\author[1]{G.~Agakishiev}
\ead{hejdar@jinr.ru}
\author[1,2]{ V.N.~Pechenov}
\ead{v.pechenov@gsi.de}
\author[1]{V.S.~Rikhvitsky}
\ead{rqvtsk@mail.ru}

\address[1]{Joint Institute for Nuclear Research, Dubna, Russia}
\address[2]{GSI Helmholtzzentrum f\"{u}r Schwerionenforschung, 64291~Darmstadt, Germany}

\begin{abstract}
The search for charged particle tracks in detectors with linear sensing elements, such as wire chambers, strip silicon detectors etc., starts with identification of straight track segments. The latter are deduced by constructing all possible combinations of activated detector elements. In the high multiplicity environment, with many activated detector elements, this causes large combinatorics and significantly reduces the performance of the track finding algorithms. In this report we trace this problem back to determination of the parameters of a straight line build from a number of skew lines (sensing elements of a detector) in space. Based on the intersection points of two hyperbolas, we demonstrate an analytic solution in the case of four skew detector elements.The procedure can also be applied as a first approximation to the more general case of curved track finding.
\end{abstract}


\end{frontmatter}


\section{Introduction}
In the following we discuss the problem on the example of the HADES experiment, but the proposed technique is more general and can be applied to any detector with the linear sensing elements.

An important part of the track finding algorithm of the HADES
spectrometer \cite{Agakishiev} is a sequential search for all
combinations of activated drift cells of mini-drift chambers (MDC). Out of four drift chambers in each 6 sectors, two
are located before and two behind the field of the toroidal magnet.

The residual magnetic field inside MDCs are rather small, which justifies the usage of straight tracks before and after the filed region, covered by two MDCs. Each chamber are composed of six sense wire layers oriented in six different stereo angles, {\emph i.e.}~$\pm 0^0$, $\pm 20^0$, $\pm 40^0$ and has around one thousand sense wires. A single track may activate up to two cells per layer yielding up to 24 activated cells in two MDCs. Moreover, depending on the energy and the size of the colliding system, a given event may have as much as 30 charged particles. The objective is to find all track candidates in a given event by building all possible combinations out of the list of activated drift cells. This approach is, however, impractical, due to a large number of combinations to be considered.

In this report we demonstrate that the performance of the track candidate search can be significantly improved by exploiting straight track segments. In doing so, we derive the parameters of a straight track passing through \emph{four} activated drift cells. The algorithm can be used for any tracking detectors with not less than four planes of linear sensing elements of two or more orientations.

The positions of sensing elements inside the detector are taken to be known. The only requirement of our method is the existence of skew linear detector elements. Furthermore, two of these elements can be parallel to each other.

We provide a method which allows to find initial approximations of
straight tracks based on the geometric problem formulated by Hermann
Schubert \cite{Schubert}; in the $n$ dimensional space the maximum
number of possibilities for $2n - 2$ lines to be intersected by the
$(n - 1)$-th line can be calculated as \cite{Kleiman,Reid,Campo}:

$$ C_n=\frac{1}{n}\cdot C^{n-1}_{2n-2}
$$

where $C_n$ is the Catalan number.

In our case we are considering intersection of four lines in the 3-dimentional space by a fifth one, i.e., $n$ = 3, which, according to formula above yields 0, 1, 2 or infinite solutions.

\begin{figure}[h]\centering
\includegraphics[width=.4 \textwidth]{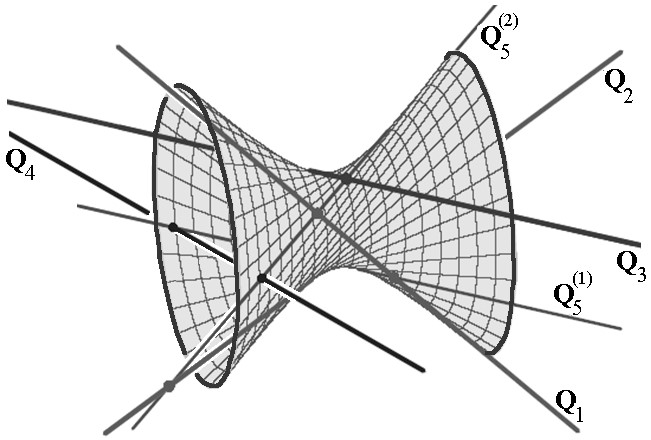}
\hspace*{2mm}
\includegraphics[width=.4 \textwidth]{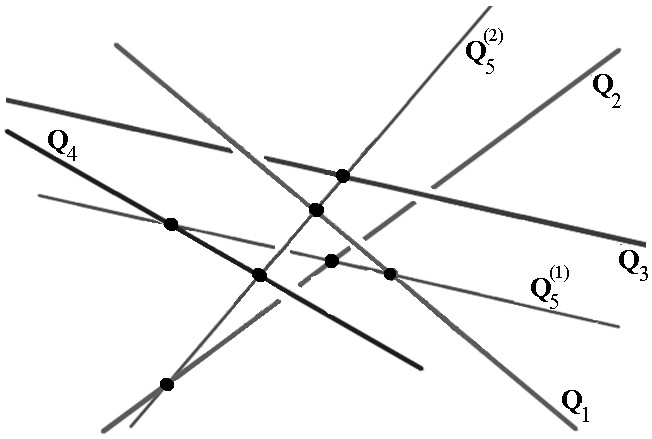}
\caption{Two spatial lines ${\bf Q}_5^{(1)}$ and ${\bf Q}_5^{(2)}$ intersect four randomly generated lines ${\bf Q}_1$, ${\bf Q}_2$, ${\bf Q}_3$ and ${\bf Q}_4$, the first three of which lie on the surface of a hyperboloid of one sheet ("Shukhov's Tower"). Note that ${\bf Q}_5^{(2)}$ and ${\bf Q}_3$ mutually intersect outside the boundaries of the figure.}\label{Fig2}
\end{figure}

\begin{figure}[h]\centering
\includegraphics[width=.4 \textwidth]{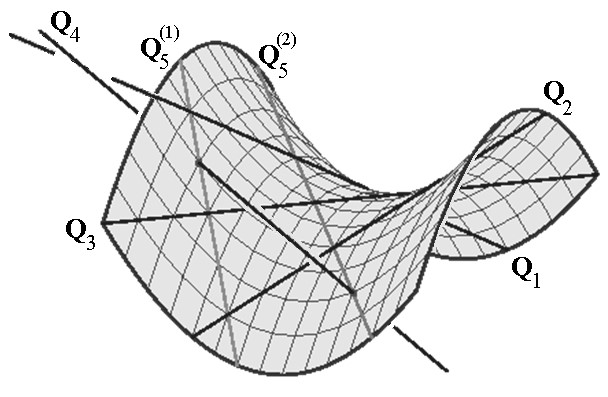}
\hspace*{2mm}
\includegraphics[width=.4 \textwidth]{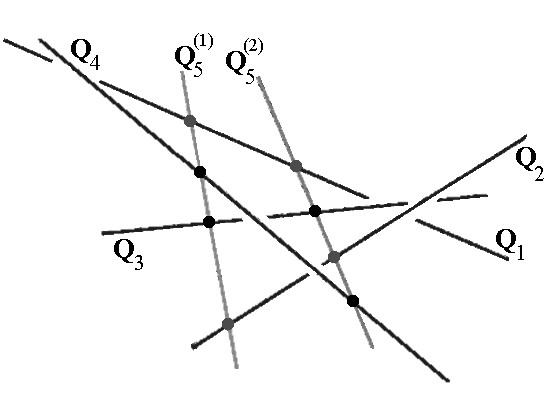}
\caption{Two spatial lines ${\bf Q}_5^{(1)}$ and ${\bf Q}_5^{(2)}$ intersect four generated lines ${\bf Q}_1$, ${\bf Q}_2$, ${\bf Q}_3$ and ${\bf Q}_4$, the first three of which lie on the surface of a hyperbolic paraboloid ("saddle").}\label{Fig3}
\end{figure}

A solution of the experimental problem of finding the explicit
equations of the fifth straight line is based on the consideration
of ruled surfaces (the �hyperboloid method�) and is considered
known. It can be illustrated by geometrical constructions
\cite{edu}. At this point we randomly generate three skew lines
${\bf Q}_1$, ${\bf Q}_2$, ${\bf Q}_3$, which uniquely lie on the
algebraic surface of second order; either on a one-sheeted
hyperboloid or hyperbolic paraboloid, which are referred to as
quadrics (cf. Figs. \ref{Fig2} and \ref{Fig3}). Furthermore there
are two families of lines on quadrics; the lines belonging to the
different families always intersect while this is not the case for
the lines form the same family. For clarity we call the family of
generated lines as \emph{first}.

The fourth straight line, ${\bf Q}_4$, intersects the quadric at 0, 1, 2 points or completely belongs to it. In the case of intersection at precisely two points, through each such point there passes one straight line of the \emph{second} family. We take them for solutions ${\bf Q}_5^{(1)}$ and ${\bf Q}_5^{(2)}$, since they must intersect the lines of the \emph{first} family ${\bf Q}_1$, ${\bf Q}_2$, ${\bf Q}_3$ and intersect ${\bf Q}_4$ by construction.

The ${\bf Q}_4$ line has, in general, either no common point with the quadric or intersects it twice; that corresponds to 0 or 2 solutions, mentioned above. However, under some circumstances the ${\bf Q}_4$ line has one common point with the quadric or completely belongs to it, which corresponds to one or infinite solutions, respectively.

In order to find out the explicit equations for the fifth line, we
consider in this contribution an alternative �hyperbolic method�,
which is of algebraic nature and was not entirely addressed in
\cite{Avramenko}.

Specifically we consider the following procedure. A charged particle which passes through the detector volume activates a number of drift wires along its path. From the list of activated wires we first randomly choose \emph{four} of them. The equations of activated wires can be easily derived using their known manufactured spatial positions. Next, we use the �hyperbolic method� to fix the explicit equation of the fifth line, i.e., initial approximation for the equations of tracks we are looking for. In fact, we find equations of two lines crossing the \emph{four} given lines. We expect that one of them is, for instance the ${\bf Q}_5^{(1)}$ line, a track candidate we are after. The second solution, for instance, ${\bf Q}_5^{(2)}$ , although it indeed exists, is not considered here, because its intersection point is well beyond the detector volume.

\section{Search for intersecting �fifth� line}

To uniquely define a line in space one needs 6 parameters (coordinates of 2 points defining the line).We iteratively locate the position of the fifth line by associating it with each set of the four skewed wires. The association is accepted when for a given configuration of skew wires we get 1 or 2 solutions (cf. the previous section). This fixes the position of the fifth line. Accordingly, we need $6 - 4 = 2$ parameters to define a common intersecting line.

The 4 lines can be defined via their parametric equations:

$$
{\bf Q}_i(t_i)={\bf V}_i\cdot t_i+{\bf P}_i, \quad i=1,\ldots, 4.
$$

where ${\bf Q}_i$ is a vector locating each point on a line $i$, ${\bf P}_i = (X_i, Y_i, Z_i)$ is a fixed point on the line $i$, ${\bf V}_i = (L_i, M_i, N_i)$ is its directing vector and $t$ is a parameter. The vectors ${\bf V}_1$, ${\bf V}_2$, ${\bf V}_3$, ${\bf V}_4$ and ${\bf P}_1$, ${\bf P}_2$, ${\bf P}_3$, ${\bf P}_4$ are considered to be fixed, with ${\bf V}_i\ne 0$ (i=1,\ldots, 4). The fifth line crossing the first four lines is also parametrized as:

\begin{equation}\label{eq1}
{\bf Q}_5(\tau)={\bf V}_5\cdot \tau+{\bf P}_5
\end{equation}

and some specific values of the parameter $\tau$ must point the points ${\bf x}_i={\bf Q}_5(\tau)$ of its intersection with the lines ${\bf Q}_i$ ($i=1,\ldots,4)$.

\begin{figure}[h]\centering
\includegraphics[width=.5 \textwidth]{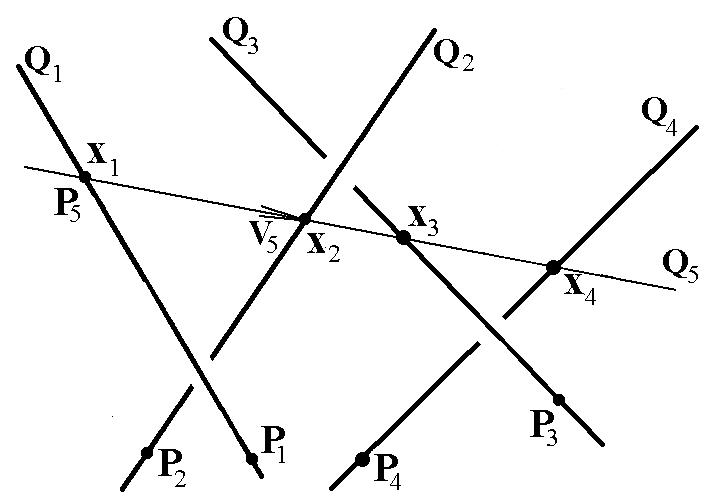}
\caption{A search for the common intersecting line $\bf Q_5.$}\label{Fig4}
\end{figure}

Method of solution. We first choose 2 arbitrary lines (\emph{principal} lines), which have 2 intersection points
$${\bf x}_1={\bf V}_1\cdot t_1+{\bf P}_1,\hspace*{2mm} {\bf x}_2={\bf V}_2\cdot t_2+{\bf P}_2.$$
 with the ${\bf Q}_5$ line. In fact, out of four lines some may have intersection points. However, the principal lines, which have to exist in a given configuration, do not intersect each other. This leads to the following expression:

$${\bf Q}_5(\tau)=({\bf x}_2-{\bf x}_1)\cdot \tau+{\bf x}_1
$$
where $\tau_1 = 0$, $\tau_2 = 1$. Using ansatz ${\bf x}_i={\bf Q}_5(t_1,t_2,\tau_i)$, $i=3,4$  (Fig. \ref{Fig4}):
\begin{equation}\label{eq2}\begin{array}{c}
{\bf Q}_5(t_1,t_2,\tau_3)={\bf Q}_3(t_3), \\
{\bf Q}_5(t_1,t_2,\tau_4)={\bf Q}_3(t_4).
\end{array}\end{equation}

Each of Eqs. (\ref{eq2}) imposes one restriction on the correspondence between $t_1$ and $t_2$. But apparently each of them introduces two new variables, in fact, each equation consists of three linear equations with respect to the new variables:
$$\left\{\begin{array}{c}
L_2t_2\tau_3-L_1t_1\tau_3+(X_2-X_1)\tau_3+L_1t_1-L_3t_3-(X_3-X_1)=0, \\
M_2t_2\tau_3-M_1t_1\tau_3+(Y_2-Y_1)\tau_3+M_1t_1-M_3t_3-(Y_3-Y_1)=0, \\
N_2t_2\tau_3-N_1t_1\tau_3+(Z_2-Z_1)\tau_3+N_1t_1-N_3t_3-(Z_3-Z_1)=0, \\
L_2t_2\tau_4-L_1t_1\tau_4+(X_2-X_1)\tau_4+L_1t_1-L_4t_4-(X_4-X_1)=0, \\
M_2t_2\tau_4-M_1t_1\tau_4+(Y_2-Y_1)\tau_4+M_1t_1-M_4t_4-(Y_4-Y_1)=0, \\
N_2t_2\tau_4-N_1t_1\tau_4+(Z_2-Z_1)\tau_4+N_1t_1-N_4t_4-(Z_4-Z_1)=0.
\end{array}\right.
$$

The result is one additional restriction in each of Eqs. (\ref{eq2}). Variables $t_3$, $t_4$, $\tau_3$ and $\tau_4$ can be easily eliminated. We ultimately get two quadratic or rather hyperbolic relations between parameters $t_1$ and $t_2$.

\begin{equation}\label{eq3}\left\{\begin{array}{c}
B_1t_1t_2+D_1t+E_1t_2+F_1=0, \\
B_2t_1t_2+D_2t+E_2t_2+F_2=0,
\end{array}\right.\end{equation}

where the coefficients (calculated via the backward substitution) are mixed vector multiplications:

\begin{equation}\label{eq4}
\begin{array}{l}
B_1=({\bf V}_2,{\bf V}_1,{\bf V}_3) \\
D_1=(({\bf P}_2-{\bf P}_3),{\bf V}_1,{\bf V}_3) \\
E_1=(({\bf P}_3-{\bf P}_1),{\bf V}_2,{\bf V}_3) \\
F_1=(({\bf P}_3-{\bf P}_1),({\bf P}_2-{\bf P}_1),{\bf V}_3)
\end{array} \quad
\begin{array}{l}
B_2=({\bf V}_2,{\bf V}_1,{\bf V}_4) \\
D_2=(({\bf P}_2-{\bf P}_4),{\bf V}_1,{\bf V}_4) \\
E_2=(({\bf P}_4-{\bf P}_1),{\bf V}_2,{\bf V}_4) \\
F_2=(({\bf P}_4-{\bf P}_1),({\bf P}_2-{\bf P}_1),{\bf V}_4)
\end{array}
\end{equation}

Eqs. (\ref{eq3}) describe two curves of the second order. In addition to Eqs. (\ref{eq3}) we will now use conventional notation of analytical geometry:
\begin{equation}\label{eq5}\left\{\begin{array}{c}
B_1xy+D_1x+E_1y+F_1=0, \\
B_2xy+D_2x+E_2y+F_2=0,
\end{array}\right.\end{equation}

We can easily find out that the curves in Eqs. (\ref{eq5}) are hyperbolas of the type $y-y_0=\pm c/ (x-x_0)$. Such hyperbolas can have 0, 1, 2 or $\infty$ intersection points (the last solution corresponds to overlapping identical hyperbolas). One can show that the set of solutions of Eqs. (\ref{eq5}) corresponds to those of H. Shubert�s problem as solved with the hyperboloid method. Consequently, we will find the intersection points of the second order curves, as defined in Eqs. (\ref{eq5}). This allows as to find the parameters $t_1$ and $t_2$, of the common intersecting line.

The system (\ref{eq5}) can be easily divided into two standard quadratic equations. At this point we introduce the following notations for both Eqs. (\ref{eq5}):
$$\begin{array}{l}
\begin{array}{cccc}
BD=\left|\begin{array}{cc}B_1 &  D_1 \\  B_2 & D_2 \end{array}\right|, &
BF=\left|\begin{array}{cc}B_1 &  F_1 \\  B_2 & F_2 \end{array}\right|, &
ED=\left|\begin{array}{cc}E_1 &  D_1 \\  E_2 & D_2 \end{array}\right|, &
EF=\left|\begin{array}{cc}E_1 &  F_1 \\  E_2 & F_2 \end{array}\right|,
\end{array} \\
\begin{array}{cccc}
BE=\left|\begin{array}{cc}B_1 &  E_1 \\  B_2 & E_2 \end{array}\right|, &
BF=\left|\begin{array}{cc}B_1 &  F_1 \\  B_2 & F_2 \end{array}\right|, &
DE=\left|\begin{array}{cc}D_1 &  E_1 \\  D_2 & E_2 \end{array}\right|, &
DF=\left|\begin{array}{cc}D_1 &  F_1 \\  D_2 & F_2 \end{array}\right|
\end{array}
\end{array}
$$

(the value of $BF$ for both equations is identical, in addition, $ED = - DE$), which yields
\begin{equation}\label{eq6}\left\{\begin{array}{c}
BD\cdot x^2+(BF+ED)\cdot x+EF=0, \\
BE\cdot y^2+(BF+DE)\cdot y +DF=0.
\end{array}\right.\end{equation}

Eqs. (\ref{eq6}) are used to determine the intersection points of hyperbolas. For each of these, in general quadratic equations, we can calculate the corresponding discriminants:
$$\begin{array}{c}
\Omega_1^2=(BF+ED)^2-4\cdot BD\cdot EF, \\
\Omega_2^2=(BF+DE)^2-4\cdot BE\cdot DF.
\end{array}\nonumber
$$

After some simplifications we deduce that $\Omega_1^2\equiv\Omega_2^2 = \Omega^2$:
\begin{eqnarray}
\Omega^2 &=&+(B_1F_2)^2-2B_1B_2F_1F_2+(B_2F_1)^2+(D_1E_2)^2-2D_1D_2E_1E_2  \nonumber \\
& & +(D_2E_1)^2 -2B_1D_1E_2F_2-2B_2D_2E_1F_1+4B_1D_2E_2F_1  \nonumber \\
& & +4B_2D_1E_1F_2 -2B_1D_2E_1F_2-2B_2D_1E_2F_1.  \nonumber
\end{eqnarray}


A common condition for the existence of real solutions to the system (\ref{eq6}) is $\Omega^2\ge 0$. In some cases the explicit boundary conditions are imposed: $BD = 0$, $BE = 0$, $BF + ED = 0$, $BF + DE = 0$, $ED = 0$ ($DE = 0$), as well as $D_i = 0$ and / or $E_i = 0$ ($i$ = 1, 2). But in real the majority of configurations are far from the boundary conditions. That yields two solutions:
\begin{eqnarray}
x_{1,2}=\frac{-BFED\pm\sqrt{\Omega^2}}{2\cdot BD},\quad y_{1,2}=\frac{-BFDE\mp\sqrt{\Omega^2}}{2\cdot BE}. \nonumber
\end{eqnarray}
(where $BFED = BF + ED$, $BFDE = BF + DE$).

In a special case, of four non-intersecting skew lines, located in parallel planes, Eq.  (\ref{eq4}) yields $B_1 = B_2 = 0$, and, moreover, if $ED\ne 0$ ($DE\ne 0$) a single solution is obtained:
$$
x_{1}=-\frac{EF}{ED},\quad y_{1}=-\frac{DF}{DE}. \nonumber
$$

All solutions of Eqs.  (\ref{eq6}), including boundary ones, can be reduced to four pairs of formulas as shown on table \ref {tab1}.
\smallskip

{
\begin{table} \centering
\begin{tabular}{|c|c|c|c|} \hline
\multicolumn{3}{|c|}{\begin{tabular}{c}  $B_1\ne 0 \vee B_2\ne 0$ \\ Two hyperbolas \\
or hyperbola and a straight line (on plane x0y) \end{tabular}} & \begin{tabular}{c}   $B_1= 0 \wedge B_2= 0$ \\ Two straight lines \\
(on plane x0y) \end{tabular} \\ \hline
$BD \ne 0 \wedge BE\ne 0$  &  $BD \ne 0 \wedge BE= 0$ & $BD = 0 \wedge BE\ne 0$ & $ED=0$  \\
& $(BFDE\ne 0)$ & $(BFDE\ne 0)$ & $(ED=-ED)$ \\
\begin{tabular}{c}  Two hyperbolas \\
with distict \\ asymptotes \\ \\ or hyperbola and \\ inclined line\end{tabular}   &
\begin{tabular}{c} Two hyperbolas \\
with coinciding \\ vertical \\ asymptotes \\ or hyperbola \\ and vertical line \end{tabular} &
\begin{tabular}{c}  Two hyperbolas \\
with coinciding \\ horizontal \\ asymptotes \\ or hyperbola \\ and horizontal line\end{tabular} &
\begin{tabular}{c}  \\ \end{tabular} \\ \hline
$\displaystyle x=\frac{-BFED\pm\sqrt{\Omega^2}}{2\cdot BD}$ & $\displaystyle x=-\frac{BF}{BD}$ & $\displaystyle x=-\frac{EF}{BFED}$ & $\displaystyle x=-\frac{EF}{ED}$ \\
$\displaystyle y=\frac{-BFDE\mp\sqrt{\Omega^2}}{2\cdot BE}$ & $\displaystyle y=-\frac{DF}{BFDE}$ &$ \displaystyle y=-\frac{BF}{BE}$ & $\displaystyle y=-\frac{DF}{DE}$ \\ \hline
\end{tabular}
\caption{Intersection points of two hyperbolas of the type $y-y_0=\pm \frac {c}{(x-x_0)}$} \label{tab1}
\end{table}
}
\smallskip

Therefore, taking the found pair of solutions (or one solution) and by putting $t_1 = x_i$, $t_2 = y_i$ ($i = 1,2$) in each pair, we get, according to (\ref{eq1}) one or two  common intersection lines:
$$
{\bf Q}_5(t_1,t_2,\tau)=[({\bf V}_2t_2+{\bf P}_2)-({\bf V}_1t_1+{\bf P}_1)]\cdot \tau+({\bf V}_1t_1+{\bf P}_1).
$$

\begin{figure*}[h]\centering
\includegraphics[width=1. \textwidth]{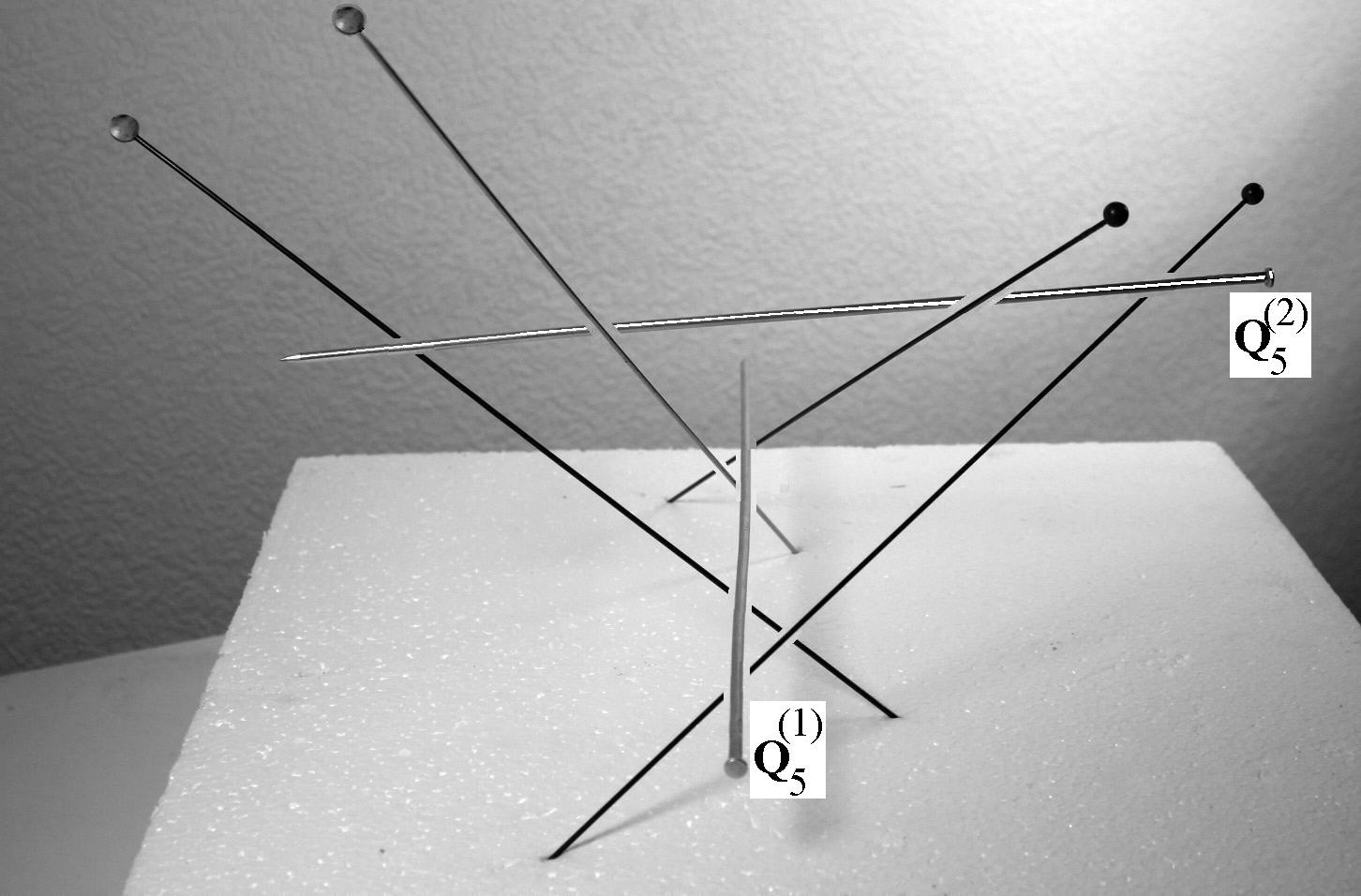}
\caption{The model of two solutions ${\bf Q}_5^{(1)}$  and ${\bf Q}_5^{(2)}$ in the form of an installation of knitting needles.}\label{Fig5}
\end{figure*}

In a real wire detector, the wires have a limited length and the only desired solution is to give a straight line that intersects the fired wires inside the detector itself. This natural condition to choose one of the two solutions is explained in
Fig.\ref{Fig5}.

Here we will find the intersection points ${\bf x}_i={\bf Q}_5(\tau_i)$ of a line with initial lines. Above we already stated that
$\tau_1 = 0$, $\tau_2 = 1$. By using backward substitutions we can get the following expressions for  $\tau_3$ and $\tau_4$:

$$
\tau_3=\frac{[({\bf P}_3-{\bf P}_5)\cdot {\bf V}_3]_x}{[{\bf V}_5\cdot {\bf V}_3]_x}=
\frac{[({\bf P}_3-{\bf P}_5)\cdot {\bf V}_3]_y}{[{\bf V}_5\cdot {\bf V}_3]_y}=
\frac{[({\bf P}_3-{\bf P}_5)\cdot {\bf V}_3]_z}{[{\bf V}_5\cdot {\bf V}_3]_z},
$$
$$
\tau_4=\frac{[({\bf P}_4-{\bf P}_5)\cdot {\bf V}_4]_x}{[{\bf V}_5\cdot {\bf V}_4]_x}=
\frac{[({\bf P}_3-{\bf P}_5)\cdot {\bf V}_4]_y}{[{\bf V}_5\cdot {\bf V}_4]_y}=
\frac{[({\bf P}_4-{\bf P}_5)\cdot {\bf V}_4]_z}{[{\bf V}_5\cdot {\bf V}_4]_z}.
$$

The equality of these proportions means the vectors $[({\bf P}_3-{\bf P}_5)\cdot {\bf V}_3]$ and $[{\bf V}_5\cdot {\bf V}_3]$, are collinear, which is a consequence of coplanarity of  $({\bf P}_3-{\bf P}_5)$, ${\bf V}_3$, and ${\bf V}_5$ vectors (the same holds true for the proportions of $\tau_4$). Disregarding the special case, $\left|[{\bf V}_5\cdot {\bf V}_3]\right|=0$ (or $\left|[{\bf V}_5\cdot {\bf V}_4]\right|=0$), which corresponds to the intersection of the 3rd (4th) line with the found 5th one at an infinitely remote point, we can take any of the presented proportions or their averages as solutions for $\tau_3$ and $\tau_4$.

We note that the fifth line, determination of which we discussed above, serves as an initial approximation for the straight track. Further improvements are possible by exploiting other activated wires along the fifth line which we plan to address in our future studies.

\section{Conclusion}

The presented method for calculating the parameters of the initial approximation for straight sections of tracks was successfully applied at the HADES facility to clean candidates in tracks obtained by the projective method from noise and the contribution from other tracks, and also to refine the initial approximation, which is used in subsequent fittings. In addition, it was used to directly search for secondary tracks after excluding the consideration of the wires from the primary tracks.

The HADES installation does not exclude situations in which the sensitive elements (in this case, four wires) can form one or two pairs of parallel straight lines. In such situations, finding a common intersecting using the previously known "hyperboloid method" is complicated; "The method of hyperbole", however, cope with them.

Note that at the HADES installation, the calculation of the initial approximation parameters for straight-line tracks using the "hyperbola method" takes a short time in the general tracking procedure.

\section*{Acknowledgements}
The authors would like to thank Valentina Kirichenko (HSE, Moscow) , E. Agakishiev, A. Jerusalimov, A. Rustamov, A. Trojan for insightful discussions and useful comments.

\section*{References}


\end{document}